\documentclass[12pt]{article}
\input{epsf.sty}
\usepackage{hyperref}
\usepackage[pdftex]{graphicx}
\usepackage{amsmath, amssymb} 
\usepackage[koi8-r]{inputenc}
\usepackage[russian, USenglish]{babel}
\newtheorem {thm}{Theorem}[section]

\newtheorem {lem}[thm]{Lemma}

\newtheorem {defn}[thm]{Definition}

\newtheorem {rem}[thm]{Remark}

\def\Cox{\hfill \Box}
\def\del{\partial}

\def\Z{{\mathbb Z}}

\def\0{{\bf 0}}

\def\ba{{\backslash}}

\def\b{\beta}

\def\e{\varepsilon}

\def\phi{\varphi}

\def\s{\sigma}

\def\o{\omega}

\def\D{\Delta}

\def\L{\Lambda}

\def\G{\Gamma}

\def\O{\Omega}

\def\T{\T}

\def\ZH{\foreignlanguage{russian}{\CYRZH}}

\def\CT{\mathfrak{CT}}
\def \atanh{\text{atanh}}

\def\PP{{\cal P}}

\def\GG{{\cal G}}

\begin{document}
\title{Gibbs-non-Gibbs properties for evolving Ising models on trees}

\author{
Aernout C.D. van Enter
\footnote{ University of Groningen, Johann Bernoulli  Institute  of Mathematics 
and Computing Science, Postbus 407, 9700  AK Groningen, The Netherlands,
\newline
 \texttt{A.C.D.v.Enter@rug.nl}, \texttt{
http://www.math.rug.nl/~aenter/ }}
\, , Victor N. Ermolaev 
\footnote{ 
University of Groningen, Johann Bernoulli  Institute  of Mathematics 
and Computing Science, Postbus 407, 9700  AK Groningen, The Netherlands
\newline
\texttt{V.N.Ermolaev@rug.nl}}
\, ,\\
Giulio Iacobelli
\footnote{ University of Groningen, Johann Bernoulli  Institute  of Mathematics 
and Computing Science, Postbus 407, 9700  AK Groningen, The Netherlands
\newline
\texttt{G.Iacobelli@rug.nl}}
\, ,  and
Christof K\"ulske
\footnote{ Ruhr-University of Bochum, Fakult\"at f\"ur Mathematik.
Postfach 102148, 44721, Bochum, Germany
\newline
 \texttt{Christof.Kuelske@ruhr-uni-bochum.de}}}
\maketitle

\begin{abstract} 
In this paper we study homogeneous Gibbs measures on  a Cayley  tree, subjected to an infinite-temperature Glauber evolution, and consider their (non-)Gibbsian properties.
We show that the intermediate Gibbs state (which in zero field is the free-boundary-condition Gibbs state) behaves different from the plus and the minus state.
E.g. at large times, all configurations are bad for the intermediate state, whereas the plus configuration never is bad for the plus state.
Moreover, we show that for each state 
there are two transitions. 
For the intermediate state there is a transition  from a Gibbsian regime to a 
non-Gibbsian regime where some, but not all 
configurations are bad, and a second  one to a regime where all 
configurations are bad. 

For the plus and minus state, the two transitions are from a Gibbsian regime 
to a non-Gibbsian one and then back to a Gibbsian regime again. 
\end{abstract}

\smallskip
\noindent {\bf AMS 2000 subject classification:}  .
\bigskip 


\section{Introduction}
In this paper we consider the Gibbsian properties of homogeneous low-temperature Ising 
Gibbs measures on trees, subjected to an infinite-temperature Glauber 
evolution.
This problem has been considered before on regular lattices see e.g. 
\cite{DR, OpKu2, KR, LR, Opoku, RRR, vanE-F-denH-R, EFHR, EKOR,  EnRu1, EnRu2}, for Ising 
spins, $n$-vector and unbounded spins, and for various finite- or 
infinite-temperature dynamics, of Glauber, Kawasaki, or diffusion type, and 
even for non-Markovian evolutions.

At high initial temperatures, or for suffiently short times,  standard
methods can be used to prove Gibbsianness, also in our situation. 
Thus the interesting case is to find out what happens 
for low initial temperatures. 
As usual (but see the mean-field analysis of \cite{ErKu}) 
low-temperature dynamics are beyond reach so far. For simplicity we will 
consider infinite-temperature dynamics, but high-temperature evolutions 
are expected to behave qualitatively similarly.
  
In contrast to what happens on regular lattices such as $\mathbb{Z}^d$, 
the Gibbsian properties of evolved Gibbs measures for models on trees turn 
out to depend on which of the different Gibbs measures (plus or minus, 
versus intermediate) one considers.  In all cases there are two transition 
times: for the intermediate measure after the first transition time it
becomes non-Gibbsian in the familiar sense  that some, but not all, 
configurations are 
`` bad''  (that is, they are points of discontinuity), while it turns out 
that after a certain later time the evolved intermediate Gibbs measure becomes 
``totally bad''; thereafter  it  has the surprising 
property that {\em all} spin configurations are discontinuity points. 

This last property is something which will not happen for the other two 
extremal invariant Gibbs measures. For those measures, although after a first 
transition time they also become non-Gibbsian, after the second transition time 
they become Gibbsian again. 

We will provide 
proofs for these results on the Cayley tree, and don't aim for the greatest 
generality here, but we will indicate why these results should be expected to 
hold more generally. We present our results both in zero and non-zero external 
fields. 

Our analysis illustrates (again)  how different 
models on trees are as compared to models on regular (amenable) lattices. 

Non-Gibbsian properties of some other  measures  of statistical 
mechanical origin on trees have been 
considered before, \cite{Hagg, HaggKu, LenyLMP}. For FK measures as well as 
fuzzy Potts models on trees, the possibility of having a positive-measure  
set of ``bad'' discontinuity points was found, and for a particular 
renormalised Ising measure the set of bad points 
was shown to have  measure zero, while having a fractal dimension.

\section{Background facts and notation}

\subsection{The Ising model on a Cayley tree}
Let $\CT(d)$ be a Cayley tree for some $d\geq 1$, that is the unique connected tree with $|\partial i|=d+1$ for all $i \in \CT(d)$.
Let $\O=\left\{-1,+1\right\}^{\CT(d)}$, endowed with the product topology. Elements in $\O$ are denoted by $\s$. A configuration $\s$
assigns to each vertex   $x\in \CT(d)$ a spin value $\s(x)=\pm1$. Denote by $\mathbb{S}$ the set of all finite subtrees of $\CT(d)$. For $\L\in\mathbb{S}$
and $\s \in \O$ we denote by $\s_{\L}$ the restriction of $\s$ to $\L$, while $\O_{\L}$ denotes the set of all such restrictions.
Let $\L \in \CT(d)$ be any set, finite or infinite. 
We denote by $E_{\L}$ its set of edges 
and by $V_{\Lambda}$ its set of vertices. 

Let now $\L \in \mathbb{S}$, hence finite. 
We will consider  the nearest-neighbour Ising model on the tree. 
The  finite-volume Gibbs measure on any finite subtree $\L$  for an Ising model 
in an inhomogeneous external field, given by  fields $h_i$ at sites $i$,  boundary condition $\omega$, at inverse temperature $\beta$,
is defined by the following Boltzmann-Gibbs distribution 
\begin{equation}
\mu^{\omega}_{\L}(\s_{\L}) = \frac{1}{Z_{\L}^{\omega}(\b, \{h_i\}_{i \in V_{\L}})} 
\exp\left\{\b \sum_{\{i,j\}\in E_{\L}} \s_i \s_j +  
\sum_{i \in V_{\L}} h_i\s_i + \sum_{\{i,j\}\atop{i \in V_{\L}, 
j \in V_{\L^c}}} \sigma_i \omega_i \right\}
\end{equation}
Infinite-volume Gibbs measures are defined by having their conditional 
probabilities of finite-volume configurations, conditioned 
on the configurations outside the volume, of this Gibbsian form, see e.g. 
\cite{Georgii} and \cite{EFS}. 
In equation form we require that for all volumes $\L$ and   configurations  
$\sigma_{\L}$ $\mu$ satisfies 
\begin{equation}
\mu(\sigma_{\L}) = \int \mu^{\omega}_{\L}(\sigma_{\L}) \mu(d \omega)
\end{equation}
The infinite-volume Gibbs measures are parametrized by the external magnetic 
fields (in most of what follows we will consider a homogeneous field $h_0$),  
and by the  inverse temperature $\b \geq 0 $. This will lead us to consider finite-volume Gibbs measures with this same  homogeneous field plus a 
possibly different boundary field.
We put $\b(1)=\infty$ and, for $d>1$,
\begin{equation}\label{hDB}
\begin{split}
&\b(d) = \text{arccoth } d = \frac12 \ln \frac{d+1}{d-1} \crcr
&h(\b, d) = \left[ d \text{ arctanh }{\left(\frac{dw - 1}{d\bar{w}-1}\right)}^{\frac12} - \text{arctanh }{\left(\frac{d - \bar{w}}{d-w}\right)}^{\frac12}\right] \mathbb{I}_{\b>\b(d)},\crcr
\end{split}
\end{equation}
where $w = \tanh \b = \bar{w}^{-1}$.

It is known (see \cite{Georgii}), that if $\b>\b(d)$ and $|h_0| \leq h(\b, d)$, then the system exibits a phase transition.
Throughout the paper we will assume $|h_0|< h(\b,d)$, $\b>\b(d)$, and $d>1$, 
whenever the opposite  is not indicated. This condition ensures the existence of three homogeneous phases $\mu^-$, $\mu^{\sharp}$, $\mu^+$.

These phases are extremal in the set of invariant infinite-volume 
Gibbs measures; 
$\mu^+$ and $\mu^-$ are also extremal in the set of all infinite-volume 
Gibbs measures, whereas 
$\mu^{\sharp}$ becomes non-extremal in this set 
below a certain temperature strictly smaller than the phase transition temperature \cite{BRZ,Iof}; however, this second transition  
will not concern us here.

Let $\G_n$ be the Cayley tree  with $n$ generations and 
$\G_{n-1}=\G_n\setminus\del\G_n$ the (sub-) Cayley tree with $n-1$ generations, where $\del\G_n$ stands for
the inner boundary of $\G_n$.
It is a known result for the Ising model on trees that the marginal on 
$\G_{n-1}$ of the finite-volume Gibbs measure on $\G_n$ is a finite-volume Gibbs measure on $\G_{n-1}$, 
with a possibly different  external magnetic field at the boundary. 
 See Appendix for how this works out in marginalizing infinite-volume Gibbs measures by using boundary laws.

 Marginalizing on $\G_{n-1}$, that is to a tree of one generation less, 
leaves us with a finite-volume Gibbs measure on $\G_{n-1}$, parametrized by the following external fields
\begin{equation}
\begin{split} \label{recu}
&i\in\del\G_{n-1}, \; h_i = h_0 + d\phi( h_n),\cr 
&i\in\G_{n-2}, \;\;\; h_i = h_0 \cr 
\end{split} 
\end{equation} 
where $\phi(x) = \atanh (\tanh \b \tanh x)$.

Thus, summarising, 
taking the marginal of an Ising model Gibbs measure 
on a tree with $n$ generations
with homogeneous boundary field $h_n$  results in an Ising model on an 
$(n-1)$-generation tree with a homogeneous boundary field $h_{n-1}$. 
The map from $h_n$ to $h_{n-1}$, \eqref{recu}, has three fixed points $h^{+},h^{\sharp}$ and 
$h^{-}$. (Equivalently, one could consider the map between the magnetisation
at generation $n$ to  the magnetisation at generation $n-1$, 
which again has the corresponding  three fixed points $m^{+}, m^{\sharp}$ and 
$m^{-}$.) Whereas $h^{+}$ and $h^{-}$ are stable, $h^{\sharp}$ is an unstable 
fixed point which implies that weak positive boundary conditions will result 
in a plus state, once one is far enough from the boundary. In other words, 
the phase transition is {\em robust} \cite {PemSt}.

These three fixed points determine the three homogeneous extremal 
invariant infinite-tree Gibbs measures mentioned above.

\subsection{Dynamics, non-Gibbsian measures, main questions}
Let $\PP(\O, \mathcal{F})$ be the set of all probability measures on $\O$ and
$\GG(\b, h_0)$ be the set of all Gibbs measures of the Ising model with an inverse temperature $\b$ and
external field $h_0$. Let $\PP_{I(B)}(\O, \mathfrak{F})$ denote the set of all $\mu \in \PP(\O, \mathcal{F})$ which are invariant under all the
graph automorphisms (translations, rotations, reflections etc). 
Let $\mu \in \GG_{I(B)}(\b, h_0)$, where $\GG_{I(B)}(\b, h_0) = \GG(\b, h_0) \bigcap \PP_{I(B)}(\O, \mathcal{F})$.

 We aim to study here the time-dependence of the Gibbsian property of the tree 
Gibbs measure $\mu^{\star}$, for $\star\in\{+,-,\sharp\}$, under an infinite-temperature Glauber dynamics. 
This is the stochastic evolution $S(t)$ which is obtained by having independent 
spin flips at each vertex at a certain given rate. In other words, we
want to investigate 
whether or not $\mu^{\star} S(t)=:\hat \mu$ is a Gibbs measure at a given time $t>0$. 

By assumption the initial measure $\mu$ is a Gibbs measure. This immediately 
guarantees the non-nullness of the measure $\mu ^{\star}S(t)$ for all $t$ (including $t=0$).
It will thus suffice to study whether the transformed measure is quasi-local or not.

Define $\hat \mu_\L(f\vert\o)=\mathbb{E}_{\hat\mu}(f|\mathcal{F}_{\L^c})(\o)$ 
to be a realization of the corresponding conditional expectation for 
bounded $f$, finite  $\L\subset \mathbb{S}$, $\o \in \O$. 
We also use the notation 
$\hat \mu_\L(f\vert\o_W)=\mathbb{E}_{\hat\mu}(f|\mathcal{F}_{W})(\o)$
when we condition only on configurations on a finite subset of sites $W\subset \L^c$.
With this notation we have e.g\
$\hat \mu_\L(f\vert\o_{\L'\backslash \L})= \int \hat\mu_\L(f\vert\o) \hat\mu(d \o_{(\L')^c})$, 
for volumes $\L'\supset \L$ where $\hat\mu(d \o_{(\L')^c})$ denotes integration over the 
variables outside of $\L'$.

The measure 
$\hat\mu$ is not quasilocal, if it is not consistent with any quasilocal 
specification. To prove this, it is enough to find a single, nonremovable, 
point of discontinuity (in the product topology) for a single $\hat\mu_\L$ for 
a single (quasi)local function $f$ \cite{RF-leHouche, EFS}. The definition of the 
non-quasilocality for the transformed measure can be refined, see in particular  \cite{RF-leHouche}. The relevant definitions read as follows:

\begin{defn}
 The measure $\hat\mu$ is not quasilocal at $\bar\eta\in \O$ if there exists $\L_0\in \mathbb{S}$ and $f$ local ( given that $\O_0$ is finite it suffices to look for $f$ local, with support $\Lambda_0$) such that no 
realization of $\hat\mu_{\Lambda_0}(f|\cdot)$ is quasilocal at $\bar\eta$.
\end{defn}

In other words, any realization of $\hat\mu_{\Lambda_0}(f|\cdot)$ must exhibit an 
essential discontinuity at $\bar\eta$; one that survives zero-measure 
modifications.
(Remember that conditional probabilities are only defined up to 
measure-zero sets.)

\begin{defn}
 For a local function $f$ as above, $\hat\mu_{\L_0}(f|\cdot)$ is $\hat\mu$-essentially discontinuous at $\bar\eta$, if there exists an $\e>0$ such that 
  \begin{equation}\label{NQC}
 \displaystyle\limsup_{\L\uparrow\infty}\sup_{\xi^1,\xi^2\atop{\L'\supset\L \atop{|\L'|<\infty}}}\vert \hat{\mu}_{\L_0}(f|\bar\eta_{\L\setminus\L_0} \xi^1_{\L'\setminus\L})-\hat{\mu}_{\L_0}(f|\bar\eta_{\L\setminus\L_0} \xi^2_{\L'\setminus\L})\vert > \e
\end{equation}
\end{defn}
If $\hat\mu_{\L_0}(f|\cdot)$ is $\hat\mu$-essentially discontinuous at $\bar\eta$, informally it means that there exists an $\e>0$ such that for every $\L\in\mathbb{S}$ there exists $\L'\supset\L$ and configurations $\xi^1, \xi^2$, such that
\begin{equation}\label{NQ1}
 \left| \hat{\mu}_{\L_0}(f|\bar\eta_{\L\setminus \L_0} \xi^1_{\L'\setminus\L} \eta)-\hat{\mu}_{\L_0}(f|\bar\eta_{\L\setminus\L_0} \xi^2_{\L'\setminus\L}\eta )\right| > \e
\end{equation}
for $\eta \in A$, where $A\in\mathcal{F}_{(\L')^{c}}$ is of positive $\hat\mu$-measure.



\begin{defn}
 $\hat\mu_{\L_0}(f|\cdot)$ is strongly discontinuous at $\bar\eta$, iff there exists an $\e>0$ such that
\begin{equation}\label{SNQC}
 \displaystyle\limsup_{\L\uparrow\infty}\sup_{\xi^1,\xi^2\atop{\L'\supset\L \atop{|\L'|<\infty}}}\inf_{{\eta^1,\eta^2}\atop{{\L''\supset\L':}\atop{|\L''|<\infty}}}\vert \hat{\mu}_{\L_0}(f|\bar\eta_{\L\setminus\L_0} \xi^1_{\L'\setminus\L}\eta^1_{\L''\setminus\L'} )-\hat{\mu}_{\L_0}(f|\bar\eta_{\L\setminus\L_0} \xi^2_{\L'\setminus\L}\eta^2_{\L''\setminus \L'} )\vert > \e
\end{equation}
\end{defn}
\begin{rem}
 Intuitively the difference is that whereas for $\hat\mu$-essential discontinuity one needs to estimate a difference on two measurable sets of positive measure, for a strong discontinuity one needs an estimate of a difference on open sets; however, because of the impossibility of conditioning on individual configurations, we get the somewhat unwieldy definitions above. 
\end{rem}

A useful tool to study whether $\hat\mu$ stays Gibbs is to consider the joint 
two-time distribution $\nu$ on $(\s, \eta)$, where the initial spins $\s$ are
distributed according to 
$\mu$, and the evolved spins $\eta$ according to $\hat\mu$. 
This joint distribution will be denoted by either $\nu$ or  ${\nu}^t$.
It can be viewed as 
a Gibbs measure on $\{-1, +1\}^{\text{\ZH}}$ with 
$\text{\ZH} = \CT(d)\cup \CT(d)$ 
consisting of two ``layers'' of $\CT(d)$. Formally, the Hamiltonian of 
${\nu}^t$ is 
\begin{equation} 
H_t(\s, \eta) = H_{\mu}(\s)-\ln p_t(\s,\eta), 
\end{equation} 
where $p_t(\s,\eta)$ is the transition kernel of the dynamics. 
We consider independent spin-flip dynamics, so 

\begin{equation} 
\ln p_t(\s,\eta) = \sum_{x\in \CT(d)}\frac12\ln\frac{1+e^{-t}}{1-e^{-t}}\s(x)\eta(x)
\end{equation} 
Let us denote
\begin{equation}\label{eq:ht}
 h^t=\frac12\ln\frac{1+e^{-t}}{1-e^{-t}}
\end{equation}
This approach to study the evolved measure as the marginal of a two-layer Gibbs
measure was introduced in \cite{vanE-F-denH-R}, and has been applied  
repeatedly since.


\begin{rem}
Here we will find for $\mu^{\sharp} S(t)$,  by making the choices $\xi_1=+1$,\\ $\xi_2= -1$, that in any open neighborhood of 
$\bar\eta$  two positive-measure sets exist, on which the limits differ, 
however, in contrast to amenable graphs, these sets are not open (which allows 
different behaviour between different evolved Gibbs measures $\mu^{\sharp}$ 
and  $\mu^+$ as regards their Gibbsianness, something which  is excluded on 
amenable graphs such as $\mathbb{Z}^d$). 
In other words we will show a $\hat\mu$-essential, although non-strong, 
discontinuity. 
\end{rem}
\bigskip
As explained in the appendix 
we have the representation of the conditional probabilities of the time-evolved measure $\mu_t$ of the form 
\begin{equation}
\begin{split}
\hat \mu_t(\eta_0| \eta_{\L \ba 0})
&=\int\mu[\eta_{\L\ba 0}](d\s_0)P_t (\s_0,\eta_0)
\end{split}
\end{equation}
with the perturbed $\eta$-dependent measure on spin configurations\\ 
$\mu[\eta_{\L\ba 0}](d\s)\equiv\mu[\eta_{\L\ba 0},\eta_0=0](d\s)$ 
whose finite-volume marginals look like 
\begin{equation}
\begin{split}
\mu[\eta_{\L'}](\s_{\L'})=C 
\exp \left\{ \b \sum_{(i,j)\in \L'}\s_i\s_j + \sum_{i\in \L'\setminus \del\L'} h_i \s_i +\sum_{i\in \del\L'} \tilde h_i \s_i\right\},
\end{split}
\end{equation}
where 
\begin{equation}
 \begin{split}\label{fields}
&   h_i=h_0+\eta_ih^t,\crcr
&  \tilde h_i=h_0+\eta_ih^t+h^\star\crcr
\end{split}
\end{equation}
where the external fields at the boundaries are given in terms of $h^\star$. This value
represents the fixed point 
of the recursion relation with homogeneous field $h_0$, \eqref{recu},
and is bijectively related with the starting measure $\mu^{\star}$. 
More generally, such a representation is always valid if the initial measure is a Markov chain 
on the tree. Markov chains can be described by boundary laws, and conditional 
probabilities of infinite-temperature time evolutions, are, for finite-volume conditionings, 
 described by boundary laws obeying recursions which are local perturbations of 
 those of the initial measure,  see the Appendix and \cite{Georgii}.

In what follows we choose $\xi^1=(+)$ and $\xi^2=(-)$. With this notation, 
for non-Gibbsianness it is enough to prove that, at $\bar\eta$, there exists an $\varepsilon>0$ 
such that, for all $\L$ there exists $\L'\supset \L$ such that 
\begin{equation}\label{lessgood}
 \left|\mu[\bar\eta_{\L\setminus0},\xi^1_{\L'\setminus\L}](\s_0)
-\mu[\bar\eta_{\L\setminus0},\xi^2_{\L'\setminus\L}](\s_0)
        \right|>\e
\end{equation}

\subsection{Marginals and \texorpdfstring{$\eta$}{eta}-dependent fields, initial field \texorpdfstring{$h_0=0$}{h0} }
To prove the non-Gibbsianness of $\hat\mu$, we will have to consider the 
phase transition behaviour of the Gibbs measures on the first layer in 
various external fields. These external fields are determined  by the various 
conditionings, as well as by the choice of the initial Gibbs measure.

%
%
Let $k,m$ be integers with $k<m$, let us denote $\L' = \G_m$ and $\L = \G_k$.
Consider first the case $h_0=0$. Marginalizing on $\G_m$ leaves us with a finite-volume Gibbs measure on $\G_m$ denoted by ${\nu}^{h^\star}_{\G_m}$ and parametrized by the following external fields
\begin{equation}
\begin{split} 
&i\in\del\G_{m}, \; h_i =\eta_i h^t+d\phi(h^\star),\cr 
&i\in\G_{m-1}, \; h_i = \eta_i h^t \cr 
\end{split} 
\end{equation} 

In order to apply the (marginalisation) procedure to the $\eta$-dependent finite-volume Gibbs measure ${\nu}^{h^\star}_{\G_m}$ on $\G_m$ we need to identify the role played by $\eta$.
It can be shown that taking the marginal on $\G_{m-1}$ of the finite-volume Gibbs measure on $\G_m$ (summing out the spin $\s\in \del\G_m$) gives us a finite-volume Gibbs measure on $\G_{m-1}$ with an external field at the boundary equal to 
\begin{equation}\label{iter}
 h_i=\eta_ih^t+\sum_{l\sim i}\phi(\eta_l h^t)
\end{equation}
Here the sum is over the nearest neighbors $l\in \G_{m}$. 

 The equation \eqref{iter} tells us how the configurations $\eta_{\del\G_m}$ will affect the field acting on $i\in \del\G_{m-1}$ after having taken one-generation marginal.

\begin{figure}[htb]
	\centering
	\includegraphics[height=8.5cm]{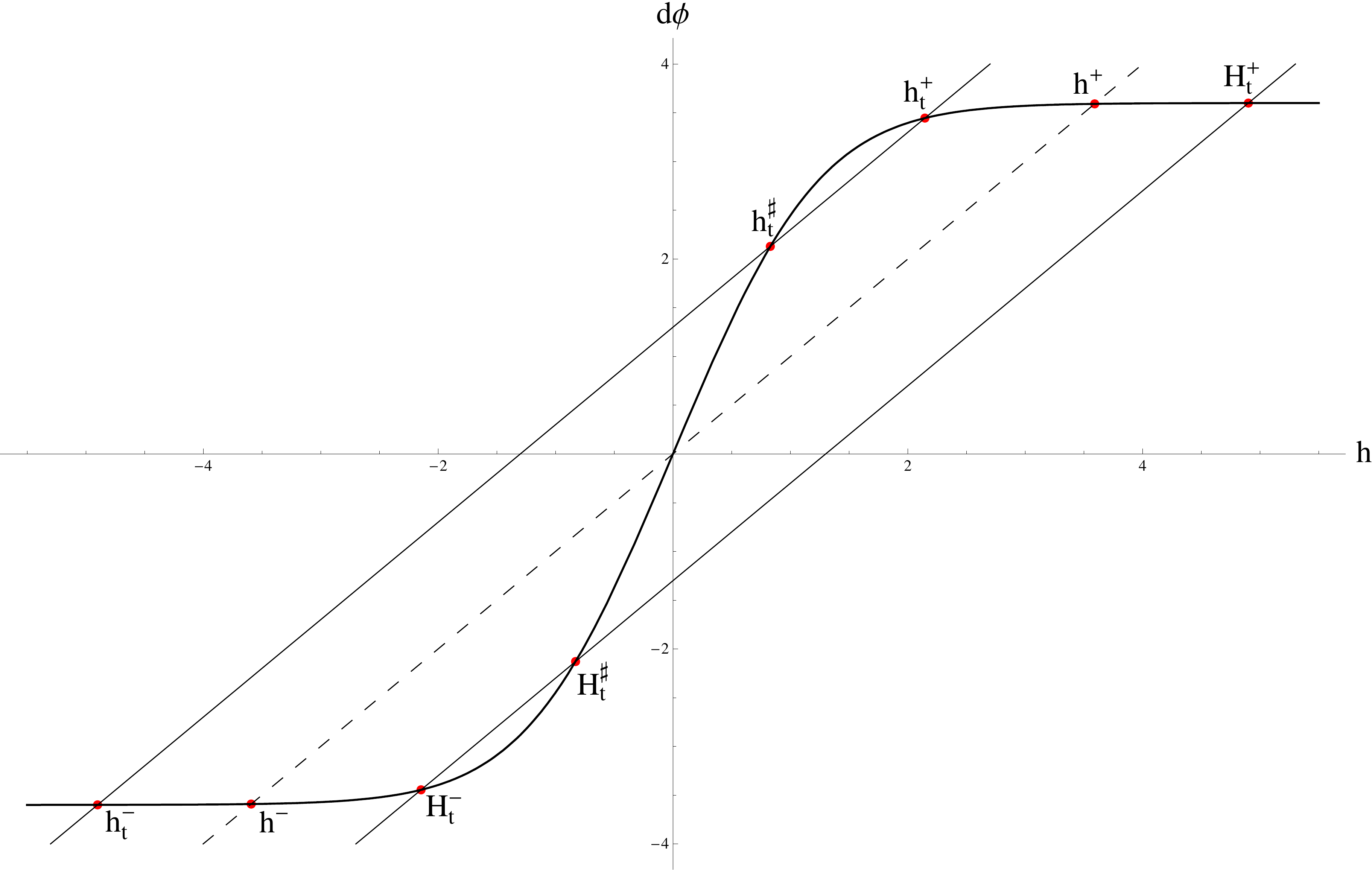}
	\caption[]{Fixed points{\protect\footnotemark}}
	\label{fixed-points}
\end{figure}

The configuration $\eta_{\G_m\setminus\G_k}$ will govern the value of the fields at $\del\G_k$, when the marginal on $\G_k$ is taken. Let us see how:

\begin{itemize}
 \item $\eta_{\G_m\setminus\G_k} = +$
	\begin{equation}
	\begin{split}\label{+}
	 &i\in\del\G_m, \; h^0_i = h^t+d\phi(h^\star),\crcr
	 &\text{after summing out the m-th generation we have}\crcr
         &i\in\del\G_{m-1}, \; h^1_i = h^t + d \phi(h^0_i), \crcr
 	 &i\in\del\G_{j},k<j<m-1, \; h^j_i = h^t + d \phi(h^{j-1}_i) \crcr
	\end{split}
	\end{equation}
	 
 \item $\eta_{\G_m\setminus\G_k} = -$
	\begin{equation}
	\begin{split}\label{-}
	 &i\in\del\G_m, \; h^0_i = -h^t+d\phi(h^\star),\crcr
	 &\text{after summing out the m-th generation we have}\crcr
         &i\in\del\G_{m-1}, \; h^1_i = -h^t + d \phi(h^0_i), \crcr
 	 &i\in\del\G_{j},k<j<m-1, \; h^j_i = -h^t + d \phi(h^{j-1}_i) \crcr
	\end{split}
	\end{equation}
\end{itemize}
Note that the above chosen $\eta$-conditioning on the annulus makes the recursion homogeneous.
Choosing $m$ big enough guarantees that the recursions \eqref{+}, \eqref{-} approach their time-dependent fixed points; we denote them respectively by $H_t^\pm$, $H_t^{\sharp}$ and $h_t^\pm$, $h_t^{\sharp}$, see Figure\eqref{fixed-points}.  

Assume that we start at time $t=0$ with the measure $\mu^{\sharp}$, then  $h^\star=h^{\sharp}=0$. It ensures that the recursions \eqref{+}, \eqref{-} will approach, respectively, $H_t^+>0$ and $h_t^-=-H_t^+<0$. $H_t^+$ represents the biggest stable fixed point for the $\eta=+$ recursion \eqref{+}, and $h_t^-$ the smallest stable fixed point for the $\eta=-$ recursion \eqref{-}. The fact that both recursions have as a starting point the unstable fixed point $h^{\sharp}=0$ guarantees that the plus conditioning will drag the field towards $H_t^+$ and the minus one towards $h_t^-$. This will not be the case for $\mu^{+}$ and $\mu^{-}$ as we will see later.

\footnotetext{``Longum est iter per praecepta, breve et efficax per exempla'', Seneca.}

The $(\eta_{\G_m\setminus\G_k}=\pm)$-dependent marginals on $\G_k$, of the  measure on $\G_m$, are finite-volume Gibbs measures parametrized by the following fields:\\
for the case $(\eta_{\G_m\setminus\G_k}=+)$
\begin{equation} 
 \begin{split}\label{R_eta+}
  &i\in\del\G_k, \; h^{+,0}_i = \eta_ih^t+d\phi(H_t^+),\cr
  &i\in \G_{k-1}, \; h^0_i=\eta_ih^t\crcr
 \end{split}
\end{equation}
and in the case $(\eta_{\G_m\setminus\G_k}=-)$
\begin{equation} 
 \begin{split}\label{R_eta-}
  &i\in\del\G_k,  \; h^{-,0}_i = \eta_ih^t+d\phi(h_t^-),\cr
  &i\in \G_{k-1}, \; h^0_i=\eta_ih^t\crcr
 \end{split}
\end{equation}

\begin{rem}
 Notice that only the fields at $\del\G_k$ depend on $\eta_{\G_m\setminus\G_k}$  and not the ones acting on the interior. We emphasize that the broadcasting is absorbed by the boundary and has no direct influence on the interior.
\end{rem}
\bigskip 

Now we investigate how the recursion relation $h^j_i = \eta_ih^t + \sum_{l\sim i} \phi(h^{j-1}_l)$, obtaining by summing out generations in $\G_k$, will depend on the fixed configurations $\eta_{\G_m\setminus\G_k}=\pm$, namely on the fields 
$H_t^+,\ h_t^-$ acting on the generation $\del\G_{k+1}$. We emphasize that the 
annulus configurations determine the starting point of the recursion.
We will also show how the aforementioned recursion relation can be bounded from below if we are coming from $\eta_{\G_m\setminus\G_k}=+$, and from above for $\eta_{\G_m\setminus\G_k}=-$. Furthermore these bounds will turn out to be uniform with respect to $\eta_{\G_k}$ 
and with respect to $j$ (number of iterations).
\begin{lem}\label{recur}
 Given the recursion relation $h^j_i = \eta_ih^t + \sum_{l\sim i} \phi(h^{j-1}_l)$ we have :\\
$h^j_i\geq h_t^+>0$, for all $i$ and $j$, if $h^0_i=H_t^+$; and $h^j_i\leq H_t^-=-h_t^+$, for all $i$ and $j$, if $h^0_i=h_t^-$. Here $h_t^+$ is the fixed point for the homogeneous recursion $h^j = -h^t + d\phi(h^{j-1})$ with $h^0=H_t^+$.
\end{lem}

{\bf Proof}: Fixed points of the discussed recursion relation are given in the picture \eqref{fixed-points}.
The proof follows by induction. Take first the case $h_i^0=H_t^+$. Naturally 
$H_t^+>h_t^+$, so $h_i^0>h_t^+$ for all $i$. If we now assume $h^j_i>h_t^+$ for all $i$, then $h^{j+1}_i = \eta_i h^t + \sum_{l\sim i} \phi(h^{j}_l)>-h^t+d\phi(h_t^+)=h_t^+$. 
The case $h_i^0=h_t^-$ follows by symmetry; the corresponding recursion relation will be bounded from above by $H_t^-$. 

$\Cox$

\section{Results: total badness of the evolved \texorpdfstring{$\mu^{\sharp}$}{muSharp}; difference between different phases}

%

%
Let $t_2$ be defined by
\begin{equation}\label{t_2}
 h^{t_2}=h(\b,d)
\end{equation}

 \begin{thm}\label{mu-hash}
  If $\s$ is distributed according to $\mu^{\sharp}$, then after time $t_2$ all configurations $\eta$ are bad configurations (points of essential discontinuity) for the transformed measure $\mu^{\sharp} S(t)$.
 \end{thm}
\begin{rem}
The main idea is as follows: 
If the plus configuration is bad (and by symmetry the same is true for 
the minus configuration), then all configurations $\bar\eta$ will be bad. 
This is because if minus boundary conditions give a minus magnetisation for 
the conditioned $\sigma$-spin at the origin, and plus boundary conditions a 
positive one, the same holds for all $\bar\eta$ (due to FKG e.g.).
So take $\bar\eta$  to  be plus. Choosing $\xi$ to be plus in a large enough 
annulus $\Lambda' \setminus  \Lambda$ and integrating the outside with 
$\mu^{\sharp}$ will lead to an effective plus boundary condition at $\Lambda$. 
The reason is that the positive magnetisation $m^+$is an attractive fixed 
point for the recursive relation, and any positively magnetised field in 
$\Lambda'$ will lead into its domain of attraction. 
The same is true for the negative magnetisation. As there are different 
magnetisations with plus and minus boundary conditions, even in the presence 
of a weak plus field (the field is plus due the $\bar\eta$ being plus), the 
choice 
of plus or minus in the annulus influences the expected magnetisation at 
the origin, however big $\Lambda$ is.
\end{rem}
\bigskip
{\bf Proof}.
The definition of $t_2$, \eqref{t_2}, will assure we are in the phase-transition regime for the transformed system  (for $t\geq t_2 $). 
Making use of Lemma\eqref{recur}, the value of $\e$ we are after, in order to prove the essential discontinuity, is given by $\e = 2\tanh(h_t^+)$. This value corresponds to taking, for the measure coming from $\eta_{\G_m\setminus\G_k}=+$, the smallest positive field along all the $k-1$ iterations, namely $h_t^+$.
The field at the origin is given by $h^k = \eta_0 h^t +(d+1)\phi(h^{k-1})$ and could be roughly bounded from below $$h^k = \eta_0 h^t +(d+1)\phi(h^{k-1})\geq -h^t+d\phi(h^+_t)=h^+_t$$
Thus the corresponding single-site measure is given by $\nu_+(\s_0)=\frac{e^{h^+_t\s_0}}{e^{h^+_t}+e^{-h^+_t}}$, so 
$$\mu[\bar\eta_{\G_k} (+)_{\G_m\setminus\G_k}](\s_0)=\tanh(h^+_t)$$ 

Analogously for the measure coming from $\eta_{\G_m\setminus\G_k}=-$,  we take the biggest negative value along all the $k-1$ iterations, that is $H_t^-=-h_t^+$, therefore 
$\nu_-(\s_0)=\frac{e^{-h^+_t\s_0}}{e^{h^+_t}+e^{-h^+_t}}$ and 
$$\mu[\bar\eta_{\G_k} (-)_{\G_m\setminus\G_k}](\s_0)=\tanh(-h^+_t)$$
For $\e = 2\tanh(h_t^+)$ the inequality \eqref{lessgood} holds.
Let us notice that $\e$ is chosen uniformly with respect to $\eta$, thanks to the uniform bounds appearing in Lemma\eqref{recur}.
This ensures the $\hat\mu$-essential discontinuity in any point.

$\Cox$


As we mentioned before, the previous argument does not hold for $\mu^+$ and $\mu^{-}$. We treat here only the $\mu^+$ case, the $\mu^-$ case is completely symmetrical. So, in case we start with the plus measure, even conditioning on a minus configuration in the annulus, due to the plus influence from the boundary will lead to a measure on $\G_k$ that looks like the plus measure in a negative field.
\begin{lem}\label{no_bad}
Given the starting measure $\mu^+$, the fields acting on $\del\G_m$ for the  
marginal measure on $\G_m$, which are given by  $h^0_i=\eta_ih^t+d\phi(h^+)$, $i\in\del\G_m$, satisfy the following inequality
\begin{equation}
 \eta_ih^t+d\phi_{\b}(h^+)>h_t^{\sharp}(d,\b)
\end{equation}
for all $d>1$, $\b>\b(d)$ and for all $t\in [t_2,\infty)$.
\end{lem}
{\bf Proof}: Let $t_2$ be as in \eqref{t_2}. We need to show that
$d\phi_{\b}(h^+(d,\b))>h_t^{\sharp}(d,\b)+h^t$ in the aforementioned region of parameters. First of all we note that the expression on the right-hand side is zero in the limit $t\uparrow\infty$, and it is a decreasing function of $t$. So in order to prove the lemma it is enough to show
\begin{equation}
 d\phi_{\b}(h^+(d,\b))>h_{t_2}^{\sharp}(d,\b)+h^{t_2}
\end{equation}
Using that $h_{t_2}^{\sharp}(d,\b)$ is a fixed point for the $(-)$ recursion at $t=t_2$, we arrive at
\begin{equation} 
 d\phi_{\b}(h^+(d,\b))>d\phi_{\b}(h_{t_2}^{\sharp}(d,\b)), 
\end{equation}
Note that $h_{t_2}^{\sharp}(d,\b)=h_c(d,\b)>0$, where $h_c(d,\b)$ is a tangent point  to $d\phi(x)$ such that $d\phi'(h_c(d,\b))=1$.
We show that $h^+>h_c(d,\b)$. In fact we know that $d\phi(h^+)-h^+=0$. Using the mean-value theorem together with the fact that $d\phi(0)=0$, we write $d\phi'(\xi)h^+-h^+=0$. It implies that $\xi$ is such that $d\phi'(\xi)=1$. Using then that $d\phi'$ is a decreasing function it follows that the domain of $\xi$, namely $(0,h^+)$ has to contain $h_c(d,\b)$; so $h^+>h_c(d,\b)$.
 Using then the monotonicity of the functions $\phi_{\b}$ the claim is proved.

$\Cox$

\begin{thm}\label{+allgood}
  If  $\s$ is distributed according to $\mu^+$, then after  time $t_2$ all configurations $\eta$ are good configurations for the transformed measure 
$\mu^+ S(t)$.
 \end{thm}
{\bf Proof}.
Based on  Lemma \eqref{no_bad}, choosing $\G_m$ big enough we make sure that the recursion relation coming from the fixed ``+''-annulus $\G_m\setminus\G_k$ will approach its fixed value $H_t^+$, so do we for the fixed ``$-$''-annulus to approach its fixed value $h_t^+$. Then the magnetic fields for the finite-volume Gibbs measure on $\G_k$ are respectively given by
\begin{equation} 
 \begin{split}
  &i\in\del\G_k, \; h^{+,0}_i = \eta_ih^t+d\phi(H_t^+),\cr
  &i\in \G_{k-1},\; h^0_i=\eta_ih^t\crcr
 \end{split}
\end{equation}
and 
\begin{equation} 
 \begin{split}
  &i\in\del\G_k, \; h^{-,0}_i = \eta_ih^t+d\phi(h_t^+),\cr
  &i\in \G_{k-1},\; h^0_i=\eta_ih^t\crcr 
 \end{split}
\end{equation}

Define $\D^j = \max_i(h_i^{+,j} - h_i^{-,j})$. This maximum is always positive, as an inductive argument shows.
We are about to prove that $\exists \delta \in (0,1)$ such that $\D^j\geq (1-\delta) \D^{j+1}$; this is equivalent to say that $\lim_{j\uparrow\infty}\D_j=0$.

\begin{equation} 
 \begin{split}
  \D^{j+1} &= \max_i(h_i^{+,j+1} - h_i^{-,j+1}) = \max_i\left[\frac{1}{d}\sum_{l\sim i}\left(d\phi(h_l^{+,j}) - d\phi(h_l^{-,j})\right)\right] \cr 
&= \max_i\left[\sum_{l\sim i}\frac{d\phi'(c_l)}{d}\left(h_l^{+,j} - h_l^{-,j}\right)\right] \leq (1-\delta) 
\max_i\left[\frac{1}{d}\sum_{l\sim i}\left(h_l^{+,j} - h_l^{-,j}\right)\right] \leq \cr
&=(1-\delta)\max_i \max_{l\sim i}\left((h_l^{+,j} - h_l^{-,j})\sum_{l\sim i}\frac{1}{d}\right)
=(1-\delta)\max_i \max_{l\sim i}\left(h_l^{+,j} - h_l^{-,j}\right)\crcr
& = (1-\delta)\D^j 
 \end{split}
\end{equation}
We used the mean-value theorem together with the fact that $d\phi'(x)<1$ for $x>h_c(d,\b)$.

$\Cox$

For $\s$ distributed according to $\mu^{\sharp}$, we will show the existence 
of an intermediate time interval, where some, but not all, configurations are 
bad for $\hat\mu$. Theorem\eqref{thm:somebad} will express this. 
We will show that the all plus and all minus configurations are good for 
$\mu^{\sharp}S(t)$ at all times in $(0,t_2)$. Moreover we will impose a condition on 
the field $h^t$ (therefore on $t$ itself), such that it guarantees the 
existence of at least one bad configuration for $\mu^{\sharp} S(t)$. 

We will find a  $t_1$, which is larger than  the minimal value of time   
for which this condition 
is satisfied. This value 
$t_1$ will turn out to be strictly less than $t_2$. This will guarantee 
that $t_1$ is small enough so that the transformed measure, conditioned on an 
all plus or  all minus $\eta$ will not exibit a 
phase transition.
\begin{rem}
Note that this implies that at the same time $t_2$ the 
intermediate state has a transition to a totally non-Gibbsian regime, where all
spin configurations are discontinuity points, whereas the plus and minus state
have a transition to a Gibbsian regime, without discontinuity points. 
\end{rem}
\bigskip
\begin{lem}\label{good:t_2}
If $\s$ is distributed according to $\mu^{\sharp}$ 
then for all $t\in(0,t_2)$ the $\eta=+$ and $\eta=-$ configurations are good configurations for the transformed measure $\mu^{\sharp} S(t)$.
\end{lem}
{\bf Proof}. As was shown before, the recursions \eqref{+}, \eqref{-} (related to the annuli) give us respectively $H_t^+$ and $h_t^-$. Let first $\eta$ be the plus configuration. In this case $h^{+,j}_i=H_t^+$ for all $i$ and $j$. In other words the field will stick to the fixed point value along the iterations. Using an inductive argument we show that $h^{-,j}_i=h^{-,j}$; that's to say that it does not depend on $i$. Based on that, it is straightforward to get a monotonicity property for $h^{-,j}$, namely that $h^{-,j+1}>h^{-,j}$ for all $j$. Indeed $h^{-,j+1}=h^t+d\phi(h^{-,j})>h^{-,j}$. The last inequality follows from the fact that $d\phi(x)>x-h^t$ for all $x\in[h_t^-,H_t^+)$, due to the chosen range of $t$. Recalling that for $t\in(0,t_2)$ the recursion relation $h^{-,j+1}=h^t+d\phi(h^{-,j})$ has only one fixed point, namely $H_t^+$, the lemma is proven for $\eta=+$. The $\eta=-$ case follows by symmetry.

$\Cox$

\begin{rem}
 The chosen range of times enables the existence of a unique fixed point for each of the recursions \eqref{+}, \eqref{-}, independently of $h^\star$. This means that the fields we obtain at $\del\G_k$ depend on the annuli, but they do not depend on the exterior $\G_m^c$. For this reason  Lemma\eqref{good:t_2} applies to $\s$'s distributed according to $\mu^+$ and $\mu^-$ too.
\end{rem}
\bigskip

For the sake of clarity, let us recall that $h^+$ indicates the positive stable fixed point for the recursion \eqref{recu} with $h_0=0$.

\begin{lem}\label{lemma:t_1}
Let $t_1$ be given by 
\begin{equation}\label{t_1}
h^{t_1}=h^+  
\end{equation}
then $t_1\in(0,t_2)$
\end{lem}
{\bf Proof}. Recalling equation \eqref{eq:ht}, the fact that $t_1$ lies in the interval $(0,t_2)$ is guaranteed by the truth of the inequality $h(d,\b)<d\phi(h^+)$, for $\b>\b(d)$ and $d>1$. Indeed
\begin{equation}
\begin{split}
 & h(d,\b)<d\atanh\left( w\left(\frac{d-\bar w}{d-w}\right)^{\frac12}\right) \cr
&=d\atanh\left( w\tanh(h_c)\right)=d\phi(h_c)\cr 
\end{split}
\end{equation}
Knowing that $h_c<h^+$, the monotonicity of the function $\phi$ concludes the proof.

$\Cox$

Define the ``alternating'' configuration $\eta^A$ to be $\eta^A_i=(-1)^n$ for $i\in\del\G_n$ and $n\in\mathbb{N}$, i.e. all vertices at each generation have the same sign different from the sign of the previous and the next generations. Naturally the configuration for which $-\eta^A_i=(-1)^n$ is also an ``alternating'' one.
Let us call $h^{\pm, j}_i$ the field at the vertex $i\in \del\G_{k-j}$ after $(j+1)$ applications of the recursion formula \eqref{iter}, starting respectively at $H^+_t$ or $h^-_t$. The particular structure of the ``alternating'' configuration makes the fields homogeneous at each generation; i.e., $h^{\pm, j}_i=h^{\pm, j}$, for all $i\in \del\G_{k-j}$.

\begin{thm}\label{thm:somebad}
If $\s$ is distributed according to $\mu^{\sharp}$, and $t_1$ is given by \eqref{t_1}, then for all $t\in [t_1,t_2)$ some, but not all, configurations $\eta$ are bad for the transformed measure $\mu^{\sharp} S(t)$.
\end{thm}
{\bf Proof}. Making use of Lemma\eqref{good:t_2}, Lemma\eqref{lemma:t_1}, to prove the theorem it is enough to find a particular configuration $\eta$ that will be bad for all $t\in [t_1,t_2)$. 
The ``alternating'' configurations will be shown to be bad for all $t\geq t_1$, in other words they transmit the influence of the annulus to the origin,
no matter how ``distant'' the annulus and the origin are.
As remarked before, $h^{{\pm},j}_i$ associated to the $\eta^A$ configurations depend only on $j$, and we call the corresponding values $h^{{\pm},j}$. 
Without loss of generality let us assume $\eta^A_i=+$, for $i\in \del\G_k$. By an inductive argument, based on the hypothesis $t\in[t_1,t_2)$ (which in terms of fields means $h^t \leq h^+$), and on the particular structure of the configuration $\eta^A$, we show that  $h^{+,j}\geq h^+$ and $h^{-,j}\leq 0$, for all $j$ even, namely for those $j$ which relate to generations at which $\eta^A$ is set to be $+$, and that $h^{-,j}\leq -h^+$ and $h^{+,j}\geq 0$ for $j$ odd. This will imply $h^{+,j}-h^{-,j}\geq h^+$ for all $j$.
Consider the case $j$ even.\\
For  $j=0$  we have:

\hspace{2cm}$h^{+,0}=H_t^{+}\geq h^+$,           \hspace{2.5cm}                                                         $h^{-,0}=h^t+d\phi(h_t^-)\leq 0$\\
Both inequalities hold,  because  $H_t^{+}$ is a decreasing function of $t$ whose lower bound is given by $h^+$.

Assuming the statement is true for $j$, let us see that it holds for $j+2$.
We focus first on $h^{+,(j+2)}$.
\begin{equation}
 h^{+,(j+2)}=h^t+d\phi(h^{+,(j+1)})=h^t+d\phi(-h^t+d\phi(h^{+,j})),
\end{equation} 
where the second equality is justified by the particular structure of the alternating configuration. Using the assumption $h^{+,j}\geq h^+$ and the monotonicity of $\phi$ we arrive at 
\begin{equation}
 h^{+,(j+2)}\geq h^t+d\phi(-h^t+d\phi(h^+))
\end{equation} 
The fact that $0\leq-h^t+h^+\leq h^+$ ensures that $d\phi(-h^t+h^+)\geq -h^t+h^+$. This concludes the proof for $h^{+,j}$.\\
For $h^{-,(j+2)}$ we have:
\begin{equation}
 h^{-,(j+2)}=h^t+d\phi(h^{-,(j+1)})=h^t+d\phi(-h^t+d\phi(h^{-,j}))
\end{equation}
Using always the assumption $h^{-,j}\leq 0$, the monotonicity of $\phi$, and the assumption $h^t\leq h^+$, which guarantees $h^t\leq d\phi(h^t)$,  we obtain
\begin{equation}
 h^{-,(j+2)}\leq h^t+d\phi(-h^t)\leq 0
\end{equation} 
The case $j$ odd is analogous.

$\Cox$
\begin{rem}
The above result also applies to the evolved plus and minus 
measures. Indeed the alternating configuration displays a strong discontinuity here, whereas the above analysis shows that for large times all configurations 
display a $\mu^{\sharp} S(t)$-essential but nonstrong discontinuity. 
Whether the $t_1$ used above is optimal in any sense is not known. We conjecture
that it may be for the intermediate state, but not for the plus or minus 
states.
\end{rem}
\bigskip

\section{Initial field \texorpdfstring{$h_0\neq 0$}{h0}}

Recall that $\vert h_0 \vert < h(d,\b)$, $\b>\b(d)$ and $d>1$; these conditions guarantee existence of three 
homogeneous phases for the original measure; we denote them, even if not fully consistent with the notation we have 
been using so far, $\mu_{h_0}^{+}$, $\mu_{h_0}^{-}$, and $\mu_{h_0}^{\sharp}$, just to emphasize their dependence on $h_0$.
We show that the previous results found for $h_0=0$ will also apply to the case $h_0\neq0$ but for different time values.
Let $t_+(h_0), t_-(h_0)$ be given by the following equations:
\begin{equation}
 \begin{split}
  &h_0 + h^{t_+} = h(d,\b), \crcr
  &h_0 - h^{t_-} = -h(d,\b) \crcr
 \end{split}
\end{equation}
Call 
\begin{equation}\label{t_2-t_3}
\begin{split}
 & t_2(h_0)=\min\left\{t_+(h_0), t_-(h_0)\right\},\crcr
 & t_3(h_0)=\max\left\{t_+(h_0), t_-(h_0)\right\}\cr
\end{split}
\end{equation}
Depending on the sign of the initial field, $t_+(h_0)$ might be either bigger or smaller then $t_-(h_0)$, as follows from \eqref{eq:ht}. Nevertheless the definitions of $t_2(h_0)$, and $t_3(h_0)$ will always assure $t_2(h_0)<t_3(h_0)$ (e.g. for $h_0<0$ the order is $t_2(h_0)=t_+<t_-=t_3(h_0)$). 

The time $t_2(h_0)$ indicates the time value for which the dynamic field $h^t$, taken in the opposite direction to $h_0$, will first reach a value which guarantees phase transition for the conditioned transformed measure. The time $t_3(h_0)$ refers to the analogous value, but for $h^t$ taken with the same sign to $h_0$.

Suppose w.l.o.g. that $h_0<0$. Note that for $h_0$ negative the magnetization corresponding to $\mu_{h_0}^\sharp$ is positive, see \cite{Georgii}, chapter 12.
For $t>t_3(h_0)$  there exist three fixed points for the $(-)$-recursion $h^{k+1}=h_0-h^t+d\phi(h^k)$, namely two stable ones $h_t^-(h_0)$, $h_t^+(h_0)$, and an unstable $h_t^{\sharp}(h_0)$. The existence of several fixed points makes the convergence to them be dependent on the starting point. In particular the recursion will take us to $h_t^+(h_0)$ iff the starting point, $h^{k=0}$, lies to the right of the unstable one, that is when $h^{k=0}>h_t^{\sharp}(h_0)$; it will take us to $h_t^-(h_0)$ iff $h^{k=0}<h_t^{\sharp}(h_0)$, and will stick to $h_t^{\sharp}(h_0)$ iff $h^{k=0}=h_t^{\sharp}(h_0)$.

Given that $t_3(h_0)>t_2(h_0)$, the assumption $t>t_3(h_0)$ ensures the existence of three fixed points also for the $(+)$-recursion $h^{k+1}=h_0+h^t+d\phi(h^k)$; they are denoted by $H_t^{\pm}(h_0)$, and $H_t^{\sharp}(h_0)$. 

Assume that we start at time $t=0$ with the measure $\mu_{h_0}^{\sharp}$, then the starting point for the $(\pm)$-recursions is $h^\star=h^{\sharp}(h_0)>0$.
However, for the chosen range of time, $t>t_3(h_0)$, it can be shown that $h^{\sharp}(h_0)$ will always lie to the right of $H_t^{\sharp}(h_0)$ and always to the left of $h_t^{\sharp}(h_0)$. So the next theorem reads:
\begin{thm}
 If $\s$ is distributed according to $\mu_{h_0}^{\sharp}$, then after time $t_3(h_0)$ all configurations $\eta$ are bad configurations for the transformed measure $\mu_{h_0}^{\sharp}S(t)$.
\end{thm}

$\Cox$

Analogously to the analysis for $h_0=0$, the former result will not hold for $\s$ distributed according to $\mu_{h_0}^\pm$. 

Two other results, obtained in the previous section, have equivalents for non-zero external field. 
\begin{lem}\label{good:t_2_h0}
If $\s$ is distributed according to $\mu_{h_0}^{\sharp}$,
then for all $t\in(0,t_2(h_0))$ the $\eta=+$ and $\eta=-$ configurations are good configurations for the transformed measure $\mu_{h_0}^{\sharp} S(t)$.
\end{lem}

$\Cox$

\begin{thm}\label{+allgood_h0}
  If  $\s$ is distributed according to $\mu_{h_0}^\pm$, then after time $t_3(h_0)$ all configurations $\eta$ are good configurations for the transformed measure $\mu_{h_0}^{\pm} S(t)$.
 \end{thm}

$\Cox$

\begin{rem}\label{goodness-shifted}
 It is worth remarking that the strict inequality $t_2(h_0)<t_3(h_0)$, always holding for $h_0\neq 0$, implies the non-emptiness of the interval of times $\left[ t_2(h_0),t_3(h_0)\right)$. A similar result to the one given in Theorem\eqref{thm:somebad} holds in the case $h_0\neq 0$, namely that for $t\in [t_2(h_0),t_3(h_0))$ some, but not all, configurations are bad.
In fact, it can be shown, for example in case $h_0<0$, that the time $t_2(h_0)$ corresponds to the time for which the plus configuration becomes bad, while for all times $t<t_3(h_0)$ the minus configuration will remain good. In case $h_0>0$, as symmetry may suggest, the time $t_2(h_0)$ will be the threshold for the minus configuration to become bad, while the plus configuration will be good till $t=t_3(h_0)$.
\end{rem}

Encouraged by the many analogies between the $h_0=0$ case and the $h_0\neq 0$ case, one might ask what one can say about the $(h_0\neq 0)$-equivalent of the time $t_1$,
\eqref{t_1}. Pursuing the former, let us define the values of times $\hat t_+$, $\hat t_-$ by the following equalities

\begin{equation}
 \begin{split}
  &h^{\hat t_+} = h_0+d\phi(h^+(h_0)) - h^\sharp(h_0), \crcr
  -&h^{\hat t_-} = h_0+d\phi(h^-(h_0)) - h^\sharp(h_0) \crcr
 \end{split}
\end{equation}
and define further
\begin{equation}\label{t_1-h0}
 t_1(h_0) = \max\left\{\hat t_+, \hat t_-\right\}
\end{equation}

\begin{figure}[!ht]
	\centering
	\includegraphics[height=8.5cm]{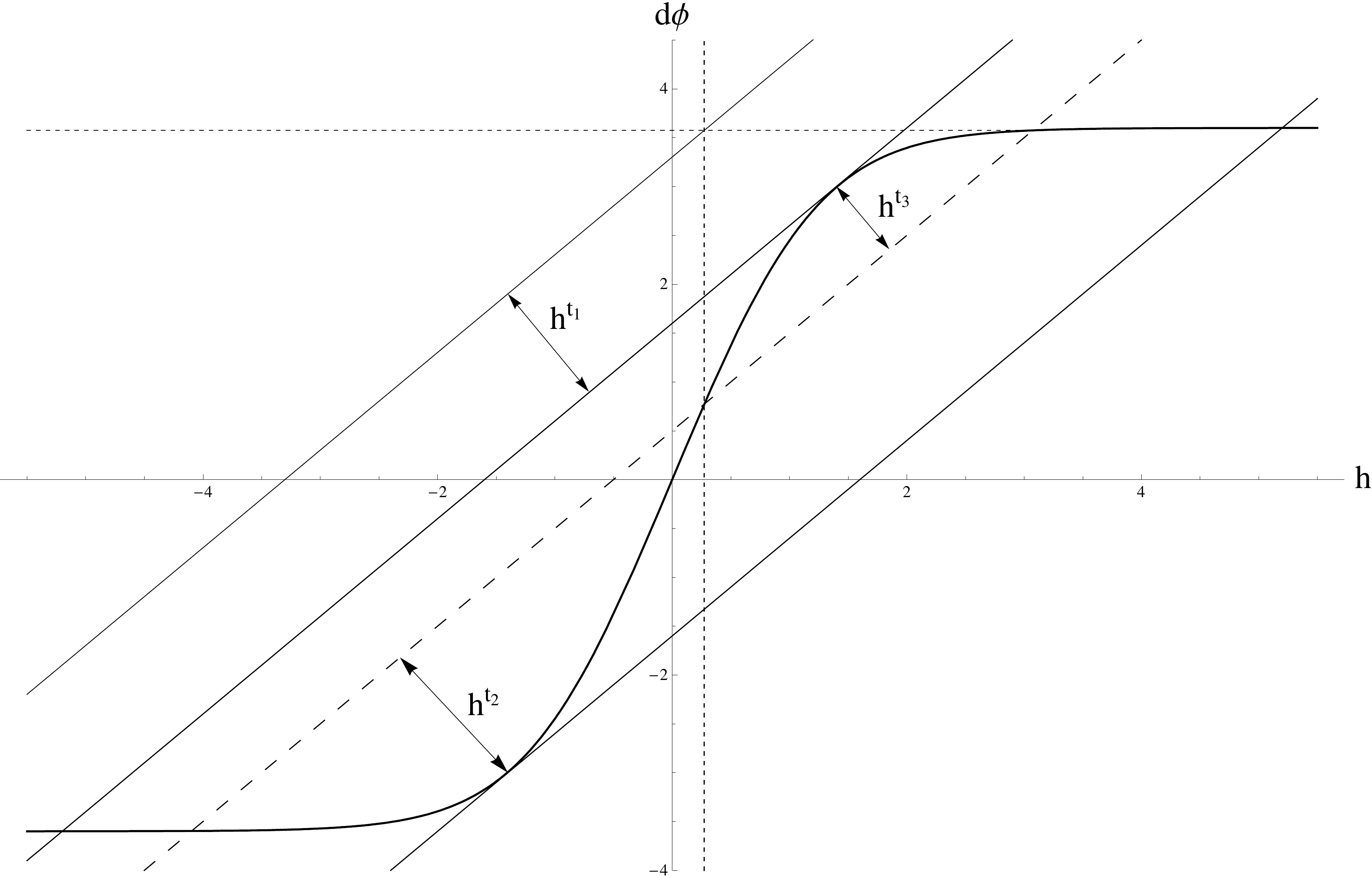}
	\caption[]{times}
	\label{times-h0}
\end{figure}

The picture \eqref{times-h0} helps to understand the role played by the different times so far defined.

It can be shown that $t_1(h_0)<t_3(h_0)$ for $\vert h_0\vert<h(d,\b)$. Nonetheless the relation between time $t_1(h_0)$ and $t_2(h_0)$ is not so trivial as we will show.
The next lemma formalizes that for all time $t\geq t_1(h_0)$ the ``alternating'' configurations are bad for $\s$ distributed according to $\mu_{h_0}^{\sharp}$.

\begin{lem}\label{good:somebad_h0}
If $\s$ is distributed according to $\mu^{\sharp}_{h_0}$, and $t_1(h_0)$ is given by \eqref{t_1-h0}, then for all $t>t_1(h_0)$ ``alternating'' configurations are bad for the transformed measure $\mu_{h_0}^{\sharp}S(t)$.
\end{lem}
{\bf Proof.}  The proof follows the same route taken in the proof of Theorem\eqref{thm:somebad} with some modifications on the bounds. Nontheless we reckon it is instructive to sketch the main points at least for $h_0<0$. For $t>t_1(h_0)$ an inductive argument leads to the following bounds:
$$\text{for even } j, \hspace{0.8cm} h^{+,j}\geq h^+(h_0) \text{ and } h^{-,j}\leq h^\sharp(h_0),\hspace{0.7cm}$$
$$\text{for odd } j, \hspace{1cm} h^{+,j}\geq -h^{\sharp}(h_0) \text{ and } h^{-,j}\leq -h^+(h_0),$$
therefore $h^{+,j}-h^{-,j}\geq h^+(h_0)-h^\sharp (h_0)$ for all $j$.

$\Cox$

The previous lemma together with Remark\eqref{goodness-shifted} shows that if $\s$ is distributed according to $\mu_{h_0}^{\sharp}$, then for all $t\in [t_1(h_0),t_3(h_0))$ some, but not all, configurations are bad. There are then two different time intervals where some, but not all, configurations are bad.
We will not leave the reader wondering
how these two intervals relate. We will show the existence of a critical value $h_0^c$ such that for $\vert h_0\vert>h_0^c$ we have $[ t_1(h_0),t_3(h_0)) \subset[t_2(h_0),t_3(h_0))$, for $\vert h_0\vert<h_0^c$ the inclusion is reversed, namely $[t_1(h_0),t_3(h_0))\supset[t_2(h_0),t_3(h_0))$, and for  $\vert h_0\vert=h_0^c$ the two intervals coincide. 
\begin{rem}
 In the small-field  regime $\vert h_0\vert<h_0^c$ we have that the 
``alternating'' configuration becomes bad before the all 
plus and the all minus configurations. In that case, the dominant effect is 
that the alternating character of the conditioning provides some cancellations, just as in the zero-field case.
 
In the other regimes we can just say what follows from $t_1(h_0)<t_3(h_0)$, i.e. that the ``alternating'' configurations become bad before the homogeneous configuration with all $\eta$'s aligned with $h_0$, that is $\eta= sign (h_0)$. The impossibility to state something more in the other regimes is due to the fact that $t_1(h_0)$ is not a ``sharp'' threshold for the ``alternating'' configurations to become bad. However, in this case having a ``bad'' configuration, one may
need to counteract the effect of the field, thus in a positive external field, the minus configuration becomes bad at an earlier time than the alternating one. 
\end{rem}

To explore the latter inclusions we need to compare the values $t_1(h_0)$ and $t_2(h_0)$, or equivalently $h^{t_1(h_0)}$ and $h^{t_2(h_0)}$.
Consider the difference between the fields
\begin{equation}
 f(h_0):= h^{t_1(h_0)}-h^{t_2(h_0)}
\end{equation}
Based on the definitions of the times, \eqref{t_1-h0}, \eqref{t_2-t_3} it turns out that the function $f$ is even. So we might focus on its behaviour only for negative values of the initial field $h_0$.
For such values of the field the function has the following form
\begin{equation}
 f(h_0)=h^{+}(h_0)-h^{\sharp}(h_0)+h_0-h(d,\b)
\end{equation}
First of all the limit values of $f$ in the interval $(-h(d,\b),0)$ are given by $$\lim_{h_0\downarrow -h(d,\b)}f(h_0)=-2h(d,\b),$$ $$\lim_{h_0\uparrow 0}f(h_0)=h^+-h(d,\b)$$
Note that the second limit value is positive, as  has been explained in the proof of Lemma\eqref{no_bad}, while the first one is negative by the definition of $h(d,\b)$, and by \eqref{hDB}.
Taking now the derivative of $f$ with respect to $h_0$ we obtain
\begin{equation}
 f'(h_0)=(h^{+}(h_0))'-(h^{\sharp}(h_0))'+1
\end{equation}
Using the only thing we know about $h^{+}(h_0)$, $h^{\sharp}(h_0)$, namely that they are fixed points for the recursion $h^{k+1}=h_0+d\phi(h^k)$, the following equalities turn out to hold $$ (h^{+}(h_0))'=\frac{1}{1-d\phi'(h^{+}(h_0))},$$ $$ (h^{\sharp}(h_0))'=\frac{1}{1-d\phi'(h^{\sharp}(h_0))}$$
Because $h^{\sharp}(h_0)<h_c(d,\b)$ and $h^{+}(h_0)>h_c(d,\b)$, the monotonicity of $d\phi'$ assures that $f'(h_0)>0$ for all $h_0\in(-h(d,\b),0)$.
Therefore the existence and uniqueness of $h_0^c$ is guaranteed by an application of the 
intermediate-value theorem.
We point out that the function $f$ is not differentiable in $h_0=0$. Indeed, being $f$ an even function and $\lim_{h_0\uparrow 0} f'(h_0)>0$ clarify the discontinuity.

We would like to remark that the case $h_0=0$ might be obtained from the previous analysis by taking the limit $h_0\uparrow 0$. Indeed $\lim_{h_0\uparrow 0}t_1(h_0)=t_1$,
$\lim_{h_0\uparrow 0}t_2(h_0)=\lim_{h_0\uparrow 0}t_3(h_0)=t_2$.

\section{Conclusion and final remarks}
We have shown that the Gibbs-non-Gibbs transition on trees has a number of 
different aspects, as compared to that on regular lattices. In particular, 
we have shown that different evolved Gibbs measures  can have different     
Gibbsian properties. For the evolved intermediate 
state there are two transitions, one from being Gibbsian to being 
``standard non-Gibbsian'' (having some, but not all configurations bad) and a 
second transition to a ``totally non-Gibbsian'' regime where {\em all} 
configurations are bad. Both these properties do not occur in the more 
familiar lattice and mean-field  situations.

For the plus and minus measure there are also two transitions, 
namely one after which the evolved measure becomes non-Gibbsian, and some, 
but not all, configurations become discontinuity points and a second 
one after which the measure becomes Gibbsian again; this is the behaviour which 
on the lattice occurs for an initial Gibbs measure in an external field.


 High-temperature dynamics should behave in a similar  way as 
infinite-tempera- ture dynamics, but although 
the proofs probably will be   messier,  qualitatively we don't expect 
anything new.

 Although we have worked out the case of Cayley trees, we expect our results 
to hold for a much wider class of trees. The instability of the fixed point 
$h^{\sharp}$ for example corresponds with the phase transition being robust, 
which is true in general for Ising models on  trees \cite{PemSt}.
Also, the property of plus boundary conditions in a not too strong minus field 
inducing a positively magnetised state, which was used in the proof that the 
plus configuration was good for the plus state  holds quite generally. 
The choice of bad configuration in the intermediate regime may be somewhat 
tree-dependent. Moreover, it seems problematical to identify  a unique  
measure $\mu^{\sharp}$ in a field (on random Galton-Watson trees for example). 

\bigskip
{\bf Acknowledgements:} The research of G.I. was supported by NWO. A.C.D.v.E.
thanks Gerhard Keller for first asking him the question whether non-Gibbsianness 
can become worse as time progresses, which triggered  this work.  

\begin{section}{Appendix: Boundary laws, beyond homogeneity}

It is the purpose of this appendix to explain the 
relation between the notion of a boundary law as it is used 
in the book by Georgii \cite{Georgii} and the one-sided simple recursions which 
are used in the paper. The notion of a boundary law is necessary 
to describe all the extremal phases (or more generally, all Markov chains on trees). 

To follow the notation used in Georgii, let us 
denote, for $i\sim j$,  by $Q_{ij}(\s_i, \s_j)= e^{\b \s_i \s_j + g_i \s_i +g_j \s_j }$ 
the transition matrix of the random field Ising model on the tree 
with Hamiltonian $-\b\sum_{\{i, j\}\in E}\s_i\s_j -\sum_{i}h_i \s_i$, where 
$g_i=h_i/(d+1)$, so the local field at each 
site has been symmetrically distributed among the edges to its neighbors.   

Every extremal Gibbs measure $\mu$ for the random field 
Ising model on the Cayley tree is a Markov chain on the tree (Theorem 12.12 of Georgii). 
To define what it means to be a Markov chain on the tree, consider an oriented bond $ij$, 
draw this bond horizontally such that $i$ lies to the left of $j$, 
and draw the tree embedded into the plane 
in such a way that there is no intersection between the tree and the axis 
crossing the oriented bond $ij$ in a perpendicular way. 
A measure $\mu$ is a Markov 
chain on the tree if conditioning on the semi-infinite 
spin configurations extending from $i$ to the left (the past)
is the same as conditioning on the spin configuration at the site $i$ alone, and this 
holds for all oriented bonds $ij$.  
Not all Markov chains are extremal Gibbs measures however, as the example of the free boundary condition Gibbs measure of the Ising model in zero field 
at sufficiently low temperatures shows. 
The meaning and importance of a boundary law lies in the following fact. 
A Markov chain on the tree always 
has a representation in terms of a boundary law $l_{ij}(a)$, $a=\pm$, that is for  
the finite-volume marginals it holds 
\begin{equation}\label{mc}
\mu[h](\s_{\L \cup \del_+ \L}) = \frac{1}{Z_{\L}(\b,h)} 
\prod_{k\in \partial_+ \L}l_{k k_\L}(\s_k) 
\prod_{\{ij\} \cap \L \neq \emptyset}Q_{ij}(\s_i, \s_j) 
\end{equation}
where $\partial_+ \L$ denotes the outer boundary of $\L$ and $k_{\L}$ is
the unique nearest neighbor of $k$ in $\L$. 
A boundary law is a function on oriented edges $ij$ which depends 
on the possible spin values. From its appearance in the last formula we see that, at any $ij$,  
it is defined only up to a multiplicative constant, not depending on the spin configuration $a$. 
Define therefore $q_{ij}=\frac{1}{2}\log \frac{l_{ij}(+)}{l_{ij}(-)}$ in the Ising case. 
This quantity has the character of 
a local field at the site $i$ and contains the 
full information about the boundary law in the Ising case. More precisely 
$q_{k k_\L}$ has the meaning of a local field acting on the 
spin $\s_k$ which has to be added to the Hamiltonian with free
boundary conditions in the volume $\L \cup \del_+ \L$ if the site $k$ is attached at the site $k_\L$. 

Assuming the validity of the last formula for the finite-volume marginals 
one arrives at a $Q$-dependent consistency (or recursion) relation 
that a boundary law has to satisfy. 
This recursion is formulated as (12.10) in Georgii; in the case of the Ising model 
with site-dependent fields  it translates equivalently into the recursion 
\begin{equation}\label{consi}
q_{ij}=\sum_{k \in \partial_+ i \backslash j}\frac{1}{2}\log 
\frac{e^{2 q_{k i} + \b + g_k + g_i }   + e^{-\b - g_k + g_i } }{e^{2 q_{k i} - \b + g_k - g_i }   + e^{\b - g_k - g_i } }
\end{equation}
Conversely, a  function $q_{ij}$ on all oriented bonds which is 
consistent in the sense of \eqref{consi}
defines a Markov chain by formula \eqref{mc} with the corresponding boundary law $l_{ij}$. 

Note that \eqref{consi} is a one-sided recursion which has no beginning and no end. 
It is interesting in the first step to look at homogeneous solutions, i.e. solutions not 
depending on the bond $ij$, but there may be also many other solutions, even in the case when 
the local magnetic field in the initial Hamiltonian is site-independent.  
In that case there can be non-homogeneous 
solutions  when there are more than one fixed points for the homogeneous recursion. 
Indeed, to construct a non-homogeneous solution one picks a site $j$ and looks to all oriented bonds $ij$ pointing 
to it, and picks values of $q_{ij}$ not at the fixed point. 
Then one defines  a boundary law by preimages for $q_b$'s for the oriented bonds $b$ 
going up to $ij$. 
In order to make sure that there 
are such preimages under all orders of iterations, the value 
has to be chosen such that it lies between a stable and an unstable fixed point. 

To see the meaning of the boundary law in a more intuitive or physical way 
let us make explicit the difference to the field which is already present in the original 
Hamiltonian.  We look at the asymmetric quantity which 
is centered at the local field for the first spin, namely 
$f_{ij}=q_{ij}- g_i d$ and note that it satisfies the equation 
\begin{equation}\begin{split}
f_{ij}
&=\sum_{k \in \partial_+ i \backslash j}\phi_{\b}(f_{k i}+h_k)
\end{split}
\end{equation}
with $\phi_{\b}(t)=\frac{1}{2}\log\frac{\cosh(t+\b)}{\cosh(t-\b)}$. 
With this variable we have 
\begin{equation}
\mu[h](\s_{\L \cup \del_+ \L}) = \frac{1}{Z_{\L}(\b,h)} 
e^{\sum_{\{ij\} \cap \L \neq \emptyset}\b \s_i \s_j + \sum_{i\in \L\cup \partial_+ \L}h_i \s_i+ \sum_{k\in \partial_+ \L} f_{k k_\L}\s_k} 
\end{equation}
So the $f_{i j}$ has the meaning of an additional boundary field at the site $i$ acting on top of 
the local fields which are present already in the Hamiltonian, when one computes the finite-volume marginals in a volume with a boundary site $i$ when $i$ is attached via the site $j$ 
to the inside of the volume.   


%

Now, let us enter in more detail the discussion on the dependence of boundary laws on a variation of 
local fields entering in the Hamiltonian. 
Suppose that a boundary law $l[h]$, not necessarily homogeneous, is given for the (not necessarily but possibly homogeneous) 
Hamiltonian with a field $h$.  Recall that, as we just explained, homogeneous fields $h$ may 
have very well inhomogeneous boundary laws. 
Let us consider the system now in the presence of  a local perturbation of the field $h+\D h$, possibly site-dependent, but bounded, i.e. $\sup_{k}|\D h_k|<\infty$.  
Any Gibbs measure $\mu[h]$ gives rise to a Gibbs measure $\mu[h+ \D h]$ which 
is related by the formula involving the local perturbation of the Hamiltonian of the form 
\begin{equation}\label{finvol}
\begin{split}
\mu[h+ \D h](\phi(\tilde \s)) &=\frac{\mu[h](\phi(\tilde \s) 
e^{\sum_{i}\D h_i \tilde \s_i}) }{\mu[h](e^{\sum_{i}\D h_i \tilde \s_i}) }
\end{split}
\end{equation}
where it is understood that integration is over $\tilde \s$. 
If the original Gibbs measure is actually a Markov chain 
described by the boundary law $l_{ij}\equiv l_{ij}[h]$, 
the perturbed measure is described by 
the boundary law  $l_{ij}[h+\D h]$ which is obtained by putting $l_{ij}[h+\D h]:=l_{ij}[h]$
for oriented bonds $ij$ in the outside of the region of the perturbation 
of the fields which are pointing towards the perturbation region. When passing with the recursion 
through the perturbation region of the local fields 
the $l_{ij}$'s obtain a dependence on the size of the 
perturbations. Then the forward iteration is 
used to obtain an assignment of $l$'s to all oriented bonds. 

Summarizing we have the following lemma. 

\begin{lem}\label{616161} Suppose that $h$ is an arbitrary external-field configuration, $\D h$ is an arbitrary
finite-volume perturbation of the external fields, and $\mu[h+\D h]$ is the measure which 
results from  a local perturbation from a Markov chain $\mu[h]$ which is described 
by a boundary law $l[h]$. 

Then $\mu[h+\D h]$ behaves in a quasilocal way (i.e. all expected values $\mu[h+\D h](\phi)$
on local spin functions $\phi$ are quasilocal functions of $\D h$) if and only if 
the boundary laws $\D h\mapsto l_{ij}[h+\D h]$, depending 
on field perturbations $\D h_k$'s for $k$ in the past of the oriented bond $ij$, 
behave in a quasilocal way, and this holds for all oriented bonds $ij$.  
\end{lem}

Here a vertex $k$ is said to be in the past of $ij$ if the path from $k$ to $j$ passes 
through $i$. Quasilocality is meant in the same way as it has been introduced 
in the context of finite-volume variations of spins, i.e. we say that $l$ depends quasilocally 
on a variation of fields iff 
\begin{equation}
\begin{split}
\lim_{\L\uparrow \Z^d}\sup_{\L': \L'\supset \L}\sup_{\D h|_{\L}=\D h'|_{\L}}
|l[ \D h|_{\L'} ]- l[ \D h'|_{\L'} ]| =0
\end{split}
\end{equation}
where the supremum is taken over perturbations 
$\D h|_{\L'},\D h'|_{\L'}  $ in the finite volume $\L'$ which look the same on $\L$. 

{\bf Proof.} The proof follows from the representation of the finite-volume 
Gibbs measures of $\mu[h+\D h]$ in terms of the boundary laws  $l_{ij}[h+\D h]$. 

$\Cox$

We note again that 
there is a one-to-one correspondence between simple directed field recursions with $d$ neighbors, 
as used in the paper, and boundary laws.  So we obtain the following corollary, which 
is used extensively in the paper. 

{\bf Corollary. } 
Suppose that $h$ is a homogeneous external field, $\D h$ is an arbitrary
finite-volume perturbation of external fields, and $\mu[h+\D h]$ is the measure which 
results from a local perturbation from either one of the homogeneous measures $\mu[h]$, 
corresponding to the plus, the minus or the unstable fixed points. 
Then the measures $\mu[h+\D h]$ behave in a non-quasilocal way on the field perturbations 
$\D h$ iff the corresponding solutions of the 
one-sided simple recursions for the effective fields 
behave in a non-quasilocal way. 

{\bf Some non-homogeneous Gibbs measures.} 
The discussion just given has consequences also 
for those Gibbs measures $\mu=\mu_{(l^b),\L}$ which are obtained by pasting boundary laws $l^b$ 
for oriented bonds $b$ of the form $k k_{\L}$ for some fixed subtree $\L$, so that \eqref{mc} 
is true for the particular volume $\L$. Then extend the boundary laws to have 
a prescription in the whole volume. 
Then the parameter region for non-quasilocal behavior of the resulting measure 
will  be the union of the parameter regions of non-Gibbsianness of the original measures 
taken over the $b$'s. 
 
{\bf Connection to Gibbs vs. non-Gibbs under time evolution.}  
Since the Gibbs properties of time-evolved Ising measures in infinite-temperature 
evolution can be expressed via quasilocality properties of $\D h\mapsto \mu[h+\D h]$,
for {\em finite-volume } $\D h$, we are left with the investigation of 
the locality properties of the boundary law iteration. 
A local variation of the image spins amounts to 
a local perturbation $\D h$ of the local fields. 
Indeed, denoting the time-evolved measure by $\hat \mu_t(d\eta)$, starting from 
the measure $\mu(d\s)$,  we have for finite $\L\ni 0$ 
the formula
\begin{equation}
\begin{split}
\hat \mu_t(\eta_0| \eta_{\L \ba 0})
&=\frac{\int\mu (d\s)P_t (\s_0,\eta_0) e^{h_t \sum_{i\in \L\ba 0}\eta_i \s_i }}{\int\mu (d\s)e^{h_t \sum_{i\in \L\ba 0}\eta_i \s_i }}\cr
&=:\int\mu[\eta_{\L\ba 0}](d\s_0)P_t (\s_0,\eta_0)
\end{split}
\end{equation}
with a measure $\mu[\eta_{\L\ba 0}](d\s)$ 
of the form $\mu[h+\D h]$ with a perturbation in the finite volume $\L\ba 0$. 
Finite-volume marginals of this  measure have a representation, according to Lemma \ref{616161},  
of the form \eqref{mc} with an $\eta$-dependent transition matrix $Q_{ij}[\eta](\s_i,\s_j)=
e^{\frac{h_t \eta_i 1_{i\in \L\ba 0}}{d+1} +\frac{h_t \eta_j 1_{j\in \L\ba 0}}{d+1} }Q_{i j}(\s_i,\s_j)$ 
where $Q_{i j}(\s_i,\s_j)$ is the transition matrix for the initial measure $\mu$, 
and an $\eta_{\L\ba 0}$-dependent boundary law $l_{ij}[\eta_{\L\ba 0}]$ which 
obeys the locally modified iterations for the boundary law described below \eqref{finvol}. 
Hence, non-Gibbsianness of time-evolved measures 
is detected by non-quasilocality of the perturbed boundary laws $l_{ij}[\eta_{\L\ba 0}]$.

A consequence of these remarks is that a time-evolved measure resulting from 
an initial Gibbs measure 
which is constructed by pasting finitely many boundary laws $l^b$ as described above,
will be non-Gibbsian at a parameter regime which is the union of 
the non-Gibbsian parameter regimes of the time-evolved Markov chains corresponding 
to $l^b$, over $b$.

\newpage 

\end{section}



\begin{thebibliography}{10}
\bibitem{BRZ}
P.~M.~Bleher, J.~Ruiz and V.~A.~Zagrebnov.
\newblock{On the purity of the limiting Gibbs state for the Ising model on the Bethe lattice}.
\newblock {\em Journal of Statistical Physics}, 79(1-2):473--482, 1995.

\bibitem{DR}
D.~Dereudre and S.~R{\oe}lly.
\newblock {Propagation of Gibbsianness for infinite-dimensional gradient
  Brownian diffusions}.
\newblock {\em Journal of Statistical Physics}, 121(3):511--551, 2005.


\bibitem{vanE-F-denH-R}
A.C.D. van Enter, R.~Fern{\'a}ndez, F.~Den~Hollander, and F.~Redig.
\newblock {Possible Loss and Recovery of Gibbsianness During the Stochastic
  Evolution of Gibbs Measures}.
\newblock {\em Communications in Mathematical Physics}, 226(1):101--130, 2002.

\bibitem{EFHR}
A.C.D. van Enter, R.~Fern{\'a}ndez, F.~Den~Hollander, and F.~Redig.
\newblock{A large-deviation view on dynamical Gibbs-non-Gibbs transitions}.
\newblock {\em Moscow Mathematical Journal, to appear}, arXiv:1005.0147, 2010. 


\bibitem{EFS}
A.C.D. van Enter, R.~Fern{\'a}ndez, and A.D. Sokal.
\newblock {Regularity properties and pathologies of position-space
  renormalization-group transformations: scope and limitations of Gibbsian
  theory}.
\newblock {\em Journal of Statistical Physics}, 72(5-6):879--1167, 1993.

\bibitem{EKOR}
A.C.D. van Enter, C.~K{\"u}lske, A.A. Opoku, and W.M. Ruszel.
\newblock {Gibbs-non-Gibbs properties for n-vector lattice and mean-field
  models}.
\newblock {\em Brazilian Journal of Probability and Statistics}, 24(2):226--255,
  2010.

\bibitem{EnRu1}
A.C.D. van Enter and W.M. Ruszel.
\newblock Loss and recovery of Gibbsianness for XY spins in  small external
  fields.
\newblock {\em Journal of Mathematical Physics}, 49(12):125208, 2008.

\bibitem{EnRu2}
A.C.D. van Enter and W.M. Ruszel.
\newblock {Gibbsianness versus Non-Gibbsianness of time-evolved planar rotor
  models}.
\newblock {\em Stochastic Processes and their Applications}, 119(6):1866--1888,
  2009.

\bibitem{ErKu}
V.N. Ermolaev and C.~K{\"u}lske.
\newblock {Low-temperature dynamics of the Curie-Weiss model: Periodic orbits,
  multiple histories and loss of Gibbsianness}. arXiv 1005.0954, 2010.

\bibitem{RF-leHouche}
R.~Fern{\'a}ndez.
\newblock {\em {Gibbsianness and non-Gibbsianness in lattice random fields,
  Les Houches summer school, Session LXXXIII, 2005, Mathematical Statistical Physics, A}}.
\newblock Elsevier, 2006.

\bibitem{Georgii}
H.O. Georgii.
\newblock {\em {Gibbs measures and phase transitions}}.
\newblock Walter de Gruyter, Berlin, ISBN 0-89925-462-4, 1988.


\bibitem{Hagg}
O.~H{\"a}ggstr{\"o}m.
\newblock {Almost sure quasilocality fails for the random-cluster model on a
  tree}.
\newblock {\em Journal of Statistical Physics}, 84(5):1351--1361, 1996.

\bibitem{HaggKu}
O.~H{\"a}ggstr{\"o}m and C.~K{\"u}lske.
\newblock {Gibbs properties of the fuzzy Potts model on trees and in mean
  field}.
\newblock {\em Markov Processes and Related Fields}, 10(3):477--506, 2004.

\bibitem{Iof}
D.~Ioffe.
\newblock{ On the extremality of the disordered state for the Ising model on the Bethe lattice}.
\newblock {\em Letters in Mathematical Physics}, 37(2):137--143, 1996.

\bibitem{OpKu2}
C.~K{\"u}lske and A.A. Opoku.
\newblock {The posterior metric and the goodness of Gibbsianness for transforms
  of Gibbs measures}.
\newblock {\em Electronic Journal of Probbability}, 13:1307--1344, 2008.

\bibitem{KR}
C.~K{\"u}lske and F.~Redig.
\newblock {Loss without recovery of Gibbsianness during diffusion of continuous
  spins}.
\newblock {\em Probability Theory and Related Fields}, 135(3):428--456, 2006.

\bibitem{LenyLMP}
A.~Le~Ny.
\newblock Fractal failure of quasilocality for a majority rule transformation
  on a tree.
\newblock {\em Letters in Mathematical Physics}, 54(1):11--24, 2000.

\bibitem{LR}
A.~Le~Ny and F.~Redig.
\newblock {Short time conservation of Gibbsianness under local stochastic
  evolutions}.
\newblock {\em Journal of Statistical Physics}, 109(5):1073--1090, 2002.

\bibitem{Opoku}
A.~A. Opoku.
\newblock {\em On Gibbs measures of transforms of lattice and mean-field
  systems}.
\newblock PhD thesis, Rijksuniversiteit Groningen, 2009.

\bibitem{PemSt}
R.~Pemantle, J.~Steif.
\newblock {Robust Phase Tramsitions for Heisenberg and other Models on General Trees}.
\newblock{\em Annals of Probability}, 27(2):876--912, 1999.   
\bibitem{RRR}
F.~Redig, S.~R{\oe}lly, and W.~Ruszel.
\newblock {Short-time Gibbsianness for infinite-dimensional diffusions with
  space-time interaction}.
\newblock {\em Journal of Statistical Physics}, 138(6):1124--1144, 2010.


\end{thebibliography}

\end{document}